\documentclass[aps,prl,twocolumn]{revtex4}
\usepackage{amssymb}
\usepackage{amsfonts}
\usepackage{amsmath}
\usepackage{graphicx}
\usepackage{hyperref}
\hypersetup{
	colorlinks=true,
	linkcolor=black,
	citecolor=blue,
	urlcolor=blue
}

\begin{document}
	
\title {Pressure Evolution of the Superconducting Order Parameter of Kagome Superconductor CsV$_3$Sb$_5$}
	
\author{Hai-yan Zuo$^{1}$, Wangju Yang$^1$, Zhi-fan Wu$^{1}$, Yu-qing Zhao$^1$, Qiuxia Chen$^1$, Weiming Lao$^1$, Yang-peng Qi$^{2}$, He-chang Lei$^3$, Hai Zi$^4$, and Cong Ren$^{1,5\ddag}$}
	
\affiliation{$^1$ School of Physics and Astronomy, Yunnan University, Kunming 650500, China}
	
\affiliation{$^2$ School of Physical Science and Technology, ShanghaiTech University, Shanghai 201210, China}
	
\affiliation{$^3$ Department of Physics and Beijing Key Laboratory of Opto-electronic Functional Materials and Micro-nano Devices, Renmin University of China, Beijing 100872, China}
	
\affiliation{$^4$ Physics Teaching and Research Section, West Yunnan University of Applied Science, Dali 671000, China}
	
\affiliation{$^5$ Yunnan Kunming 650500, China}

\begin{abstract}
Kagome superconductors \(\rm{AV_3Sb_5}\) (A = K, Rb, Cs) offer an outstanding platform for exploring the interplay among charge-density-wave (CDW) order, superconductivity (SC), and nontrivial topological electronic states. For CsV\(_3\)Sb\(_5\), moderate hydrostatic pressure induces a double-dome superconducting phase diagram while monotonically suppressing the CDW phase, which signals an unconventional coupling between SC and CDW. Nevertheless, the microscopic nature of the pressure-modulated superconducting state and its relationship with CDW order remain poorly understood. Here we employ soft directional point-contact spectroscopy (SDPCS), combined with superconducting resistive measurements, to probe the  superconducting order parameter of CsV\(_3\)Sb\(_5\) under hydrostatic compression. Our anisotropic PCAR measurements resolve two distinct $s$-wave superconducting gaps along the $c$-axis and a strongly anisotropic gap within the $ab$ plane. Two-band Blonder–Tinkham–Klapwijk (BTK) fits to the $c$-axis spectra yield a large gap satisfying \(2\Delta_l/k_BT_c>3.52\) and a small gap with \(2\Delta_s/k_BT_c<3.52\). The spectral-weight-averaged coupling ratio \(2\Delta_\mathrm{avg}/k_BT_c\) is enhanced in the pressure window between \(P_1\) and \(P_2\), suggesting that pressure-induced modifications of the low-energy electronic structure may contribute to the evolution of superconductivity.  For $ab$-plane junctions, the superconducting gap anisotropy is gradually suppressed above $P_1$, accompanied by the emergence of a pronounced peak-dip-hump structure suggestive of enhanced coupling between superconducting quasiparticles and a bosonic excitation. These results shew new light on the intricate entanglement between CDW and SC in kagome superconductivity.

\end{abstract}
	
\maketitle
	
\textbf{Introduction:} The coexistence and competition between superconductivity and density-wave orders are frequently observed in unconventional superconductors, including cuprates, iron-based compounds, and heavy-fermion systems, and have thus become a central theme in condensed matter physics. The kagome lattice, composed of a corner-sharing triangular network, intrinsically hosts a suite of distinctive electronic features: flat bands with quenched kinetic energy \cite{flatband1,flatband2}, van Hove singularities in proximity to the Fermi level \cite{vonHove0,vonHove1,vonHove2,vonHove3}, and nontrivial topological band invariants \cite{topolg1,topolg2}.  Recently discovered superconductivity in kagome compounds AV$_3$Sb$_5$ (A = K, Rb, Cs) adds a new physical dimension of electronic order to this novel system \cite{OrtizPRM2019,OrtizPRM2021,LeiHCCPL2021}. Intriguingly, these V-based superconductors exhibit highly unusual charge density wave (CDW) order, within which many exotic properties, such as giant anomalous Hall effect \cite{GiantHall1,GiantHall2,GiantHall3}, the time reversal symmetry breaking \cite{timeRev1,timeRev2}, and electronic nematicity \cite{nemat1,nemat2,nemat3}, have been discovered.
	
Moreover, on the side of superconducting state, the CDW order was intertwined with the superconductivity in an unconventional manner. Application of hydrostatic pressure ($P$) or chemical doping can suppress the CDW order, but enhance the superconductivity. For example, the superconductivity exhibits a superconducting double-dome in a specific pressure window between $P_1\sim 0.70$ and $P_2\sim 2.0$ GPa in CsV$_3$Sb$_5$, while the CDW order is monotonically suppressed by pressure, indicating an unusual competition between the superconductivity and the CDW order \cite{Pressure1,Pessure2,Pressure3}.  In addition, the Ti-V and Sn-Sb substitution in CsV$_3$Sb$_5$, were found to have two superconducting domes and a completely suppressed CDW transition \cite{TiSub1,TiSub2,SnSub1,SnSub2}, similar to the case of pressure-tuned superconductivity.  This usually manifests as a superconducting dome neighbouring a density-wave order in the phase diagram, and density-wave fluctuations are considered as the major pairing glue for such kagome superconductors. Under such circumstances, further careful study on the pressure-dependent superconductivity, particularly its gap structure, is highly desirable to clarify the origin of this double-dome feature and intertwinement of CDW and superconductivity in CsV$_3$Sb$_5$ \cite{ZiHPRB2024,LuXSC2023,ChenXHSB2023}.

\begin{figure}
\includegraphics[scale=0.32]{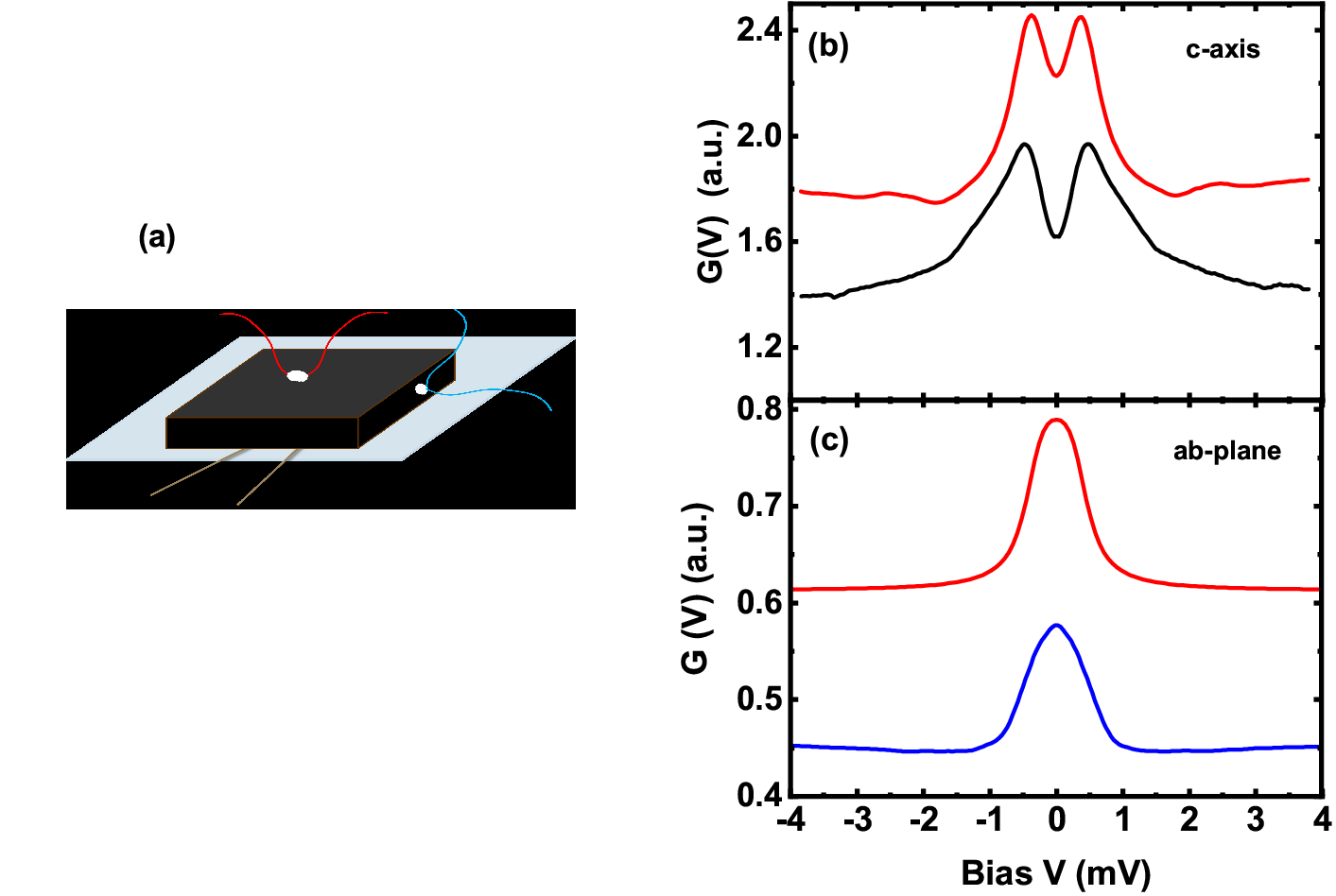}
\caption{(a) Schematic of the ``soft'' directional point-contact configuration. The black block represents the CsV$_3$Sb$_5$ single crystal, while the red and blue lines indicate the injection paths for soft point-contact Andreev reflection measurements with the current applied perpendicular ($c$-axis junction) and parallel ($ab$-plane junction) to the $ab$-plane, respectively. The typical soft point-contact spectra at $T=0.30 $ K under ambient pressure for (b) $c$-axis junction (the sensing current $I$ perpendicular to $ab$ plane) and (c) $ab$-plane junction ($I$ parallel to $ab$ plane). }
\end{figure}
	
\begin{figure}
\includegraphics[scale=0.30]{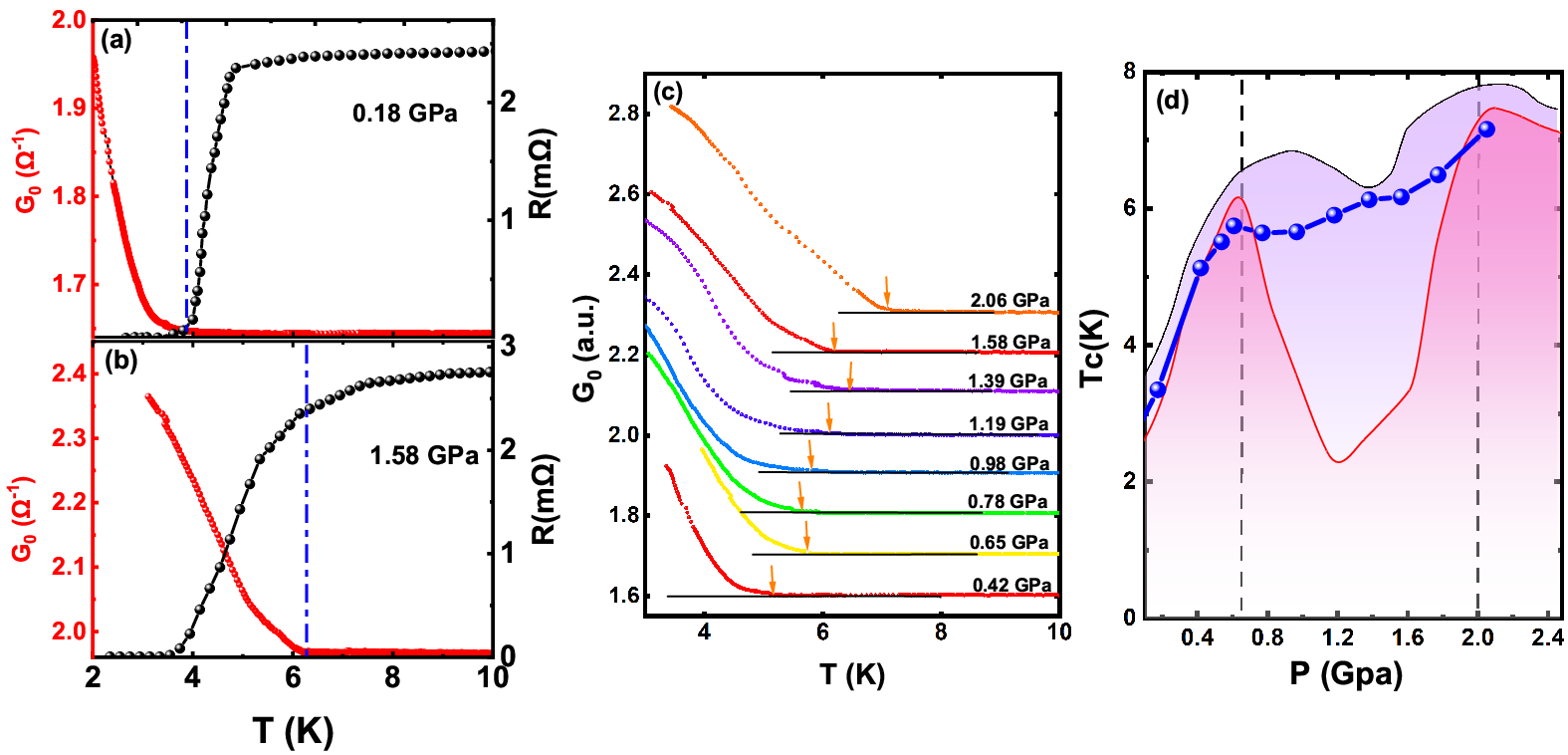}
\caption{(Color online) Comparison between the $T$-dependence of zero-bias conductance (ZBC) $G_0$ and electrical resistance of the CsV$_3$Sb$_5$ crystal under (a) $P=0.18$ GPa and (b) $P=1.58$ GPa, respectively. (c) The pressure evolution of the $T$-dependent zero bias conductance for SPCS of the CsV$_3$Sb$_5$ $c$-axis junction.  The red arrows mark the Andreev temperatures $T^A_c$. The $G_0$ curves are shifted upwards for clarity.  (d) Pressure evolution of Andreev temperature $T^A_c$ of SPCS of CsV$_3$Sb$_5$ in $T_c - P$ phase diagram (color-filled region).  The $T_c-P$ phase diagram has been constructed by electrical transport measurement adapted from ref. [28]. }
\end{figure}
	
Hydrostatic pressure serves as a clean, continuous tuning parameter that modifies lattice geometry, orbital overlap, and electronic band structure without introducing chemical disorder, thereby offering unique access to the phase diagram of correlated kagome metals. In contrast to chemical doping, which inevitably induces random site substitution and disorder scattering, pressure modulates the kagome lattice bond lengths and angles, directly shifting the energy of flat bands and van Hove singularities relative to the Fermi surface and altering the electronic susceptibility toward CDW and superconducting instabilities. Regarding high-pressure superconductivity, several critical open questions persist. It remains unclear whether the enhancement of superconductivity arises from the melting of the CDW gap at the Fermi surface, the pressure-driven shift of van Hove singularities across Fermi surface, modification of electron-phonon coupling, or the emergence of new correlated states in the CDW-quenched regime \cite{ChenXHSB2023,LuXSC2023}.
	
In this work, we conduct a directional point-contact Andreev reflection spectroscopy in kagome CsV$_3$Sb$_5$ crystals combined electrical transport measurements on CsVSb crystals under hydrostatic pressure up to 2.1 GPa. Highly reliable and reproducible Andreev reflection spectroscopy data allow us to resolve the pressure evolution of superconducting gaps, and thus the superconducting coupling strength $2\Delta/k_BT_c$.
	
\begin{figure*}
\includegraphics[scale=0.60]{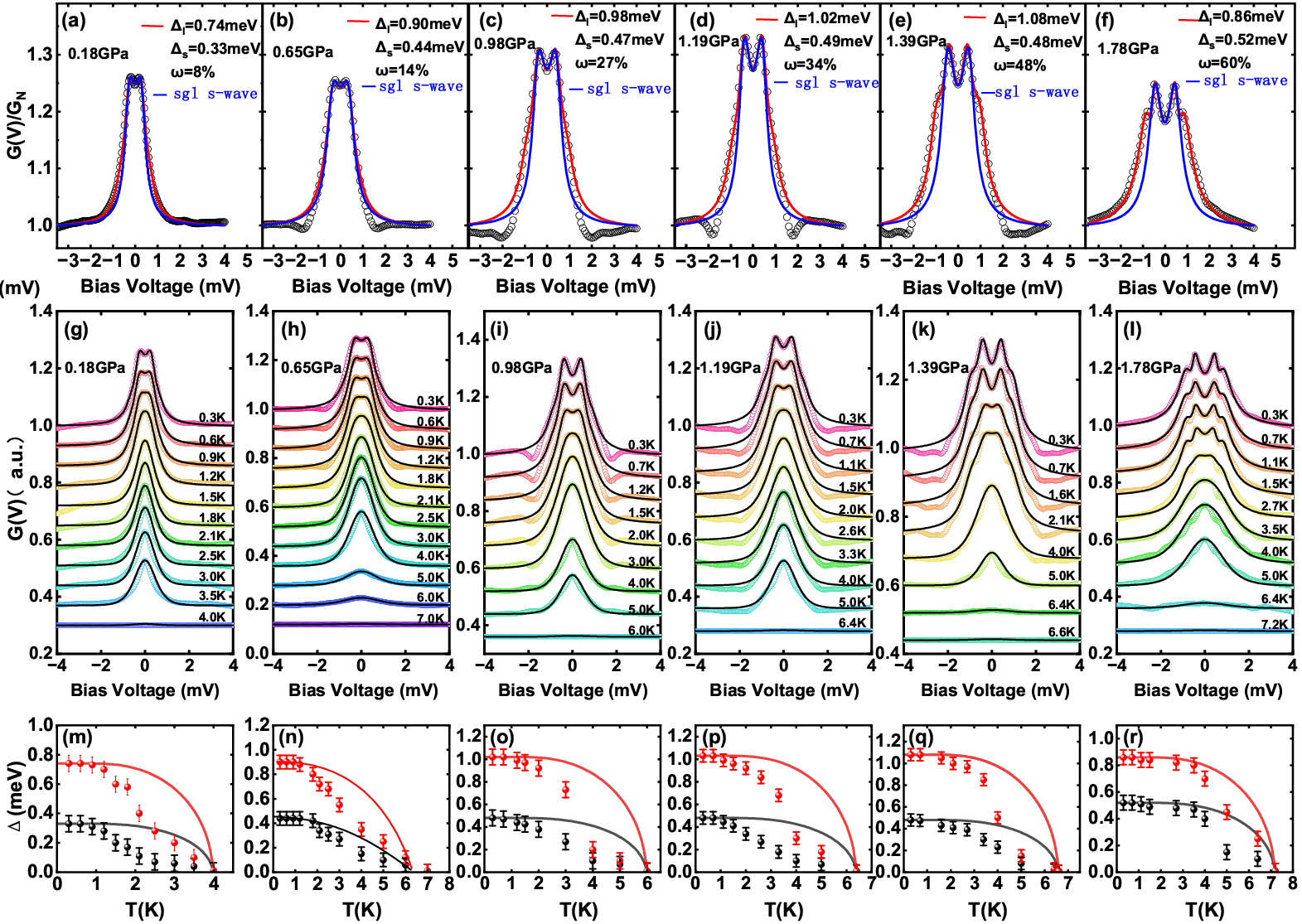}
\caption{(Color online) (a)-(f) The normalized SPCAR spectra $G(V)/G_N$ of $c$-axis CsV$_3$Sb$_5$/Ag contact at the lowest $T=0.30$ K at selected pressures.  The red solid curves are their corresponding two $s$-wave BTK-model fits.  The blue solid curves are the corresponding BTK fits with a single $s$-wave model for comparison.  (g)-(l) The $T$-evolution of the normalized Andreev conductance spectra and their fits.  All curves are shifted downward for clarity. (m)-(r) The corresponding SC gaps $\Delta$ extracted from the BTK fitting as a function of $T$.  The colored solid lines are the BCS-like temperature dependence: $\Delta(T)=\Delta(0)\tanh(1.76 \sqrt{T_c/T -1})$.}
\end{figure*}

\begin{figure}
\includegraphics[scale=0.52]{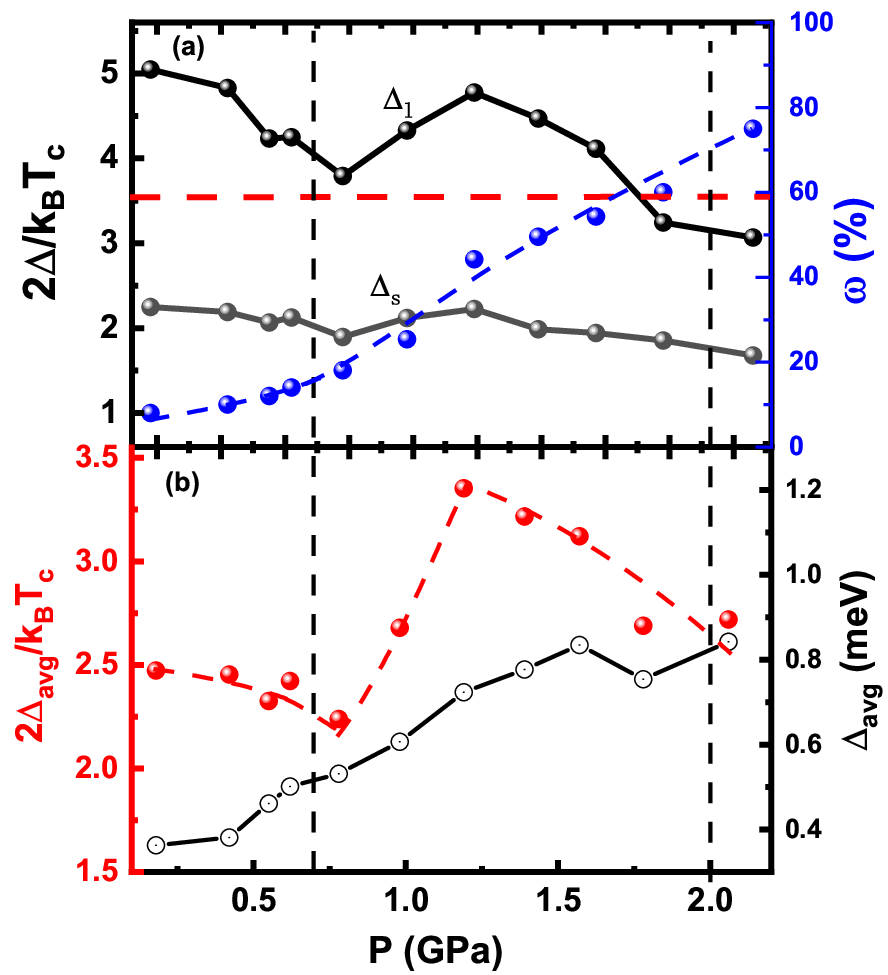}
\caption{(Color online) (a) Evolution of the superconducting coupling ratio  $2\Delta/k_BT_c$ and the spectral weight of the large gap $\omega$ as a function of pressure for $\Delta_l$ and $\Delta_s$, respectively.  The blue dashed line is the eye-guide line for the spectral weight $\omega$. The red dashed line represent the standard BCS weak-coupling strength of 3.52.   (b) The spectral weight-averaged SC gap $\Delta_{avg}$ and its coupling strength $\Delta_{avg}/k_BT_c$ as a function of pressure.  The weight-averaged SC gap $2\Delta_{avg}$ is defined as $\Delta_{avg}=\omega \Delta_l + (1-\omega)\Delta_s$.  The red solid line is the guide for eye. The black dashed lines mark two critical pressure $P_1=0.7$ GPa and $P_2\simeq 2.0$ GPa, characteristic of the double-dome in the $T_c-P$ phase diagram of CsV$_3$Sb$_5$. }
\end{figure}

\textbf{Experiment details:} Directional point-contact spectroscopy (PCS) measurements were implemented on high-quality CsV$_3$Sb$_5$ single crystals via a soft-contact experimental protocol \cite{ItalyPC}. The investigated single crystals possessed standard geometric dimensions of $3.5\times 3.0\times (0.3-0.5)$ mm$^3$ along the $a$, $b$, and $c$ crystallographic axes, respectively. Prior to measurements, the crystal surfaces were processed by cleavage and mechanical polishing to expose well-defined crystallographic planes for directional contact fabrication. Platinum wires bonded with silver paint were subsequently mounted on the freshly prepared crystal surfaces. After drying, the silver nanoparticles embedded in the paint form abundant parallel nanoscale contact arrays on the sample surface, constructing effective heterogeneous point-contact junctions with a lateral size of 50–80 $\mu$m, which function equivalently to conventional metallic-tip point contacts.
	
Two types of directional contact geometries were fabricated to realize anisotropic PCS characterization. The first category corresponds to out-of-plane $c$-axis junctions, where metallic contacts were directly fabricated on the pristine or cleaved $ab$-plane surfaces of the crystals. The second category refers to in-plane $ab$-plane junctions, which were constructed on $ac$/$bc$ surfaces perpendicular to the intrinsic $ab$-plane. For the preparation of $ab$-plane contacts, the crystals were precisely cut along the direction perpendicular to the $ab$-plane, followed by fine mechanical polishing with silicon carbide abrasive paper to obtain microscopically smooth surfaces. The structural configurations of the two types of point-contact junctions are illustrated in the schematic diagram of Fig. 1(a).
	
All differential conductance spectra $G(V) \equiv dI/dV$ were collected using a quasi-four-terminal measurement setup combined with phase-locked amplifier detection technology. The overall contact resistance of the fabricated junctions was primarily governed by the interfacial coupling between silver nanoparticles and the CsV$_3$Sb$_5$ crystal surface, falling in the range of $0.5 \Omega - 10 \Omega$. This resistance range satisfies the characteristic criteria for valid ballistic point contacts \cite{BTK2}. Representative PCS spectra measured at $T=0.30$ K for $c$-axis and $ab$-plane contact junctions are presented in Fig. 1(c) and Fig. 1(d), respectively.
	
Piston cylinder cell (PCC) was used to generate hydrostatic pressure up to 2.1 GPa with  Daphne 7373 oil the pressure transmitting medium. The pressure value in PCC was determined from the superconducting transition of Pb stripe \cite{ZiHPRB2024}. The high-pressure resistance and SPCS conductance spectrum $G(V)$ of CsV$_3$Sb$_5$ single crystal were measured in a $^3$He cryogenics.
	
\textbf{Results and Discussions:} Empirically, repeated measurements are required for pressure-tunable point-contact spectroscopy to guarantee reliable and reproducible results, since this experimental method combines two technically challenging techniques: high-pressure and point-contact spectroscopy measurements. In our prior work on the same CsV$_3$Sb$_5$ crystal, we observed an excess SPC conductance spectrum $G(V)$ at ambient pressure for $T > T_c^R$, where $T_c^R$ denotes the superconducting resistive transition temperature. This excess $G(V)$ manifests as a zero-bias conductance (ZBC) $G_0$ tail in the regime $T>T_c^R$ \cite{HeMCPRB2022,LuXPRB2021}. Such a feature introduces ambiguities in determining the intrinsic $T_c$ and performing background normalization of the resulting PCAR conductance curves. Nevertheless, with applied pressure of 0.21 GPa, the excess conductance spectrum is fully suppressed, and the resistive transition temperature \(T_\text{c}^R\) coincides with the Andreev temperature \(T_\text{c}^A\) \cite{HeMCPRB2022}.
	
\begin{figure*}
\includegraphics[scale=0.62]{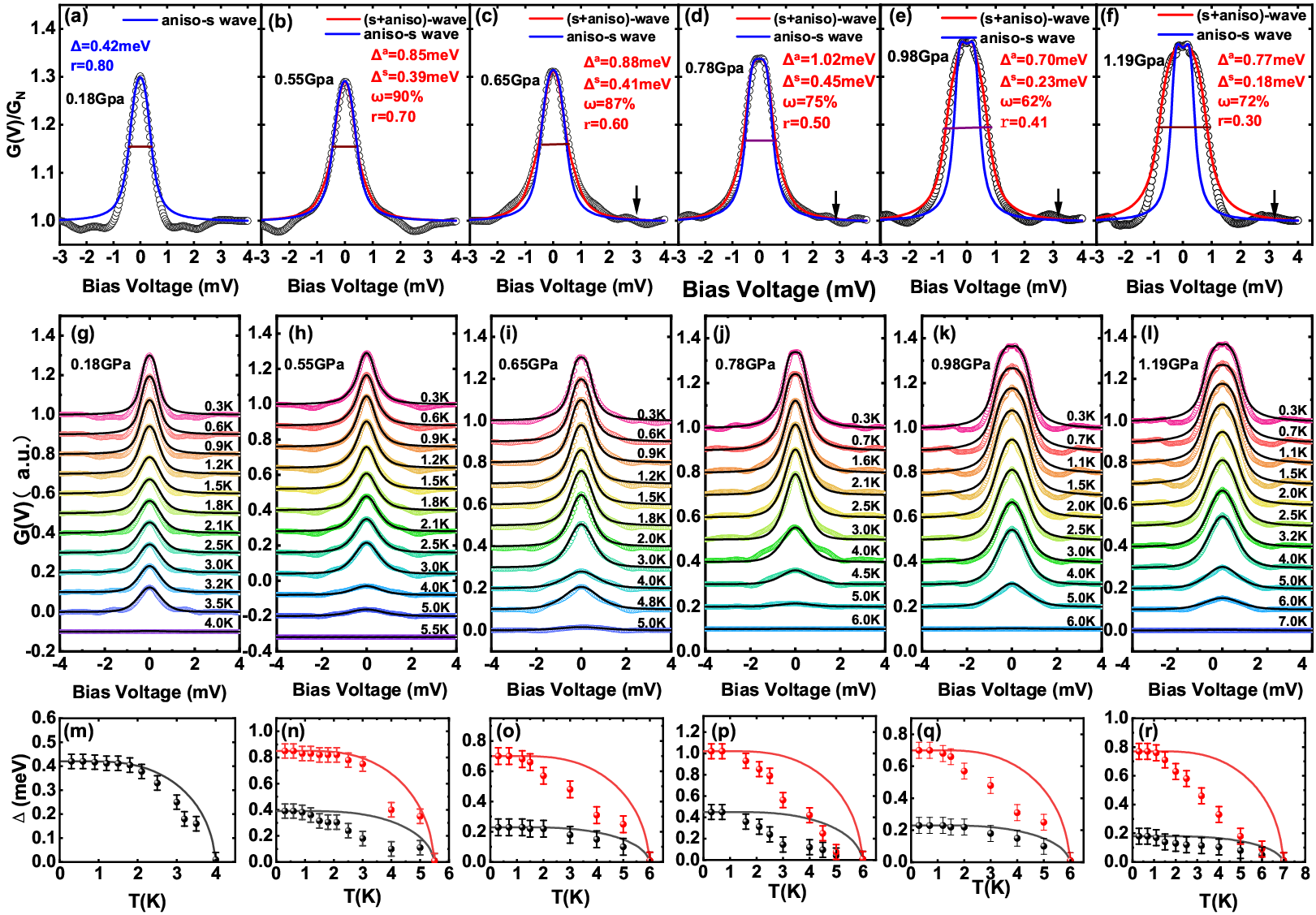}
\caption{(a)-(f) The normalized SPCAR spectra $G(V)/G_N$ of $ab$-plane CsV$_3$Sb$_5$/Ag contact at the lowest $T=0.30$ K at selected pressures.  The red solid curves are their two-gap BTK-model fits, including one full $s$-wave $G^s(V)$ and the another anisotropic $s$-wave $G^a(V)$ gap.  The blue solid curves are the corresponding anisotropic BTK fits, see text.  (g)-(l) The $T$-evolution of the normalized Andreev conductance spectra and their fits.  All PCAR spectra and their corresponding fits are shifted downward for clarity. (m)-(r) The corresponding SC gaps $\Delta$ extracted from the BTK fitting as a function of $T$.  The colored solid lines are the expected BCS gap formula: $\Delta(T)=\Delta(0)\tanh(1.76 \sqrt{T_c/T -1})$.}
\end{figure*}
	
Likewise, we fully reproduce this phenomenon in our present measurements, as shown in Fig. 1(a).  The disappearance of the ZBC $G_0$ tail and the coincidence between $T_c^R$ and $T_c^A$ under $P\simeq 0.18$ GPa and a relatively high pressure of 1.57 GPa, as illustrated in Figs. 2(a) and 2(b).  The reproducibility of point-contact spectra under pressure validates the fidelity of our experimental observations, and the elimination of both the above-$T_c^R$ excess conductance spectra enables us to probe the intrinsic superconducting gap structure of pressurized CsV$_3$Sb$_5$.
	
Figure 2(c) presents the $T$-dependent ZBC $G_0$ acquired via SPCS on CsV$_3$Sb$_5$ under varying hydrostatic pressures. The red arrows mark the Andreev temperature $T_c^A$, which characterizes the gap opening/closing temperature, termed as Andreev temperature $T_c^A$.  Using the pressure evolution of $T_c^A$, we construct the $T_c^A-P$ phase diagram displayed in Fig. 2(d), in comparison with $T_c^R-P$ phase diagram constructed from electrical transport measurements \cite{Pressure1,ChenXHSB2023}. It is worth noting that $T_c^A$ does not exhibit a dome-like behavior. The $T_c^A$, instead, in the pressure region $(P_1,P_2)$ shows a gradual enhancement without any obvious drop, in contrast to the M-shaped double-dome behavior well-established in the resistive $T_c-P$ phase diagram. A similar $P$-dependence of $T_c^A$ has been reported for pressurized CsV$_3$Sb$_5$ crystals in a previous work \cite{LuXSC2023}.  This discrepancy suggests that the pressure-induced double-dome behavior of the bulk resistive transition does not directly relates to a corresponding nonmonotonic evolution of the superconducting gap-opening temperature.
	
Figure 3 summarizes the primary point-contact Andreev reflection (PCAR) spectroscopic results obtained from $c$-axis CsV$_3$Sb$_5$/Ag junctions under hydrostatic pressures ranging from 0.18 to 2.1 GPa. Figure 3(a) displays the normalized PCAR conductance $G(V)/G_N$ measured at $T=0.30$ K, which is well fitted by a two-band $s$-wave Blonder–Tinkham–Klapwijk (BTK) model of the form $G(V)=\omega G_l(V)+(1-\omega)G_s(V)$. For direct comparison, a conventional single-gap $s$-wave BTK fitting curve is also presented. Evidently, the two-gap model provides a substantially better description of the spectral features than the single-gap model, particularly in reproducing the conductance structure around the gap edges, demonstrating the retention of dual $s$-wave superconducting gaps under pressures. The optimal two-gap fitting yields a large gap $\Delta_l \simeq 0.78$ meV, a small gap $\Delta_s \simeq 0.43$ meV, and a reduced spectral weight $\omega \simeq 8\%$ for the large-gap component.
	
Upon increasing the pressure to 0.42 GPa, the PCAR spectral features remain nearly unchanged, with both the gap magnitudes and the spectral weight of the large gap staying almost identical to those of low $P$ [Fig. 3(b)]. In contrast, when the pressure rises to 0.65 GPa — close to the critical pressure $P_1\simeq 0.70$ GPa — both $\Delta_l$ and $\Delta_s$ exhibit clear enhancement, while the spectral weight of the large gap increases substantially to 14\% (Fig. 3(c)). Further pressurization to 0.98 GPa continuously strengthens the large-gap contribution, which reaches 27\% and 60\% at 1.78 GPa.  Remarkably, all measured $c$-axis PCAR spectra across the entire pressure regime can be consistently well reproduced by the two-gap $s$-wave BTK model.
	
The reliability of the aforementioned BTK fitting procedure is further verified via $T$-dependent $c$-axis PCAR conductance measurements. Figure 3(g)–(k) presents the evolution of normalized conductance spectra $G(V)$ with temperature, and the corresponding $T$-dependent gap magnitudes at different pressures are summarized in Figs. 3(o)–(r). By extracting the gap amplitudes and superconducting transition temperatures $T_c^A$ from systematic BTK fittings, we can quantitatively evaluate the superconducting coupling strength of pressurized CsV$_3$Sb$_5$. The pressure evolution of the dimensionless coupling ratio $2\Delta/k_BT_c$ for both $\Delta_l$ and $\Delta_s$ is plotted in Fig. 4(a). Notably, the relation $2\Delta_l/k_BT_c > 3.52 > 2\Delta_s/k_BT_c$ is well satisfied, which quantitatively resembles the behavior of the canonical two-band $s$-wave superconductor MgB$_2$ \cite{ItalyPC,MgB2,MgB3}.
	
To characterize the overall gap size of the multiband superconducting state, we define a spectral-weight-averaged gap as: $\Delta_\mathrm{avg}=\omega\Delta_l+(1-\omega)\Delta_s$.  Here, $\omega$ represents the relative spectral weight of the large-gap contribution obtained from the BTK fitting \cite{HeMCPRB2022}. As displayed in Fig. 4(b), $\Delta_\mathrm{avg}$ monotonically increases with applied pressure. In contrast, the corresponding ratio $2\Delta_\mathrm{avg}/k_BT_c$ exhibits a mild downward trend with increasing pressure. Upon approaching the critical pressure $P_1\simeq 0.70$ GPa, $2\Delta_\mathrm{avg}/k_BT_c$ first rises and then declines, yielding a dome-like profile of the averaged superconducting coupling strength between $P_1$ and $P_2$. Intriguingly, the $2\Delta_\mathrm{avg}/k_BT_c$ value reaches its maximum at $P\simeq 1.19$ GPa, which coincides with the maximal difference $\delta T_c$ observed in the constructed $T$–$P$ phase diagram.
	
Next, we systematically study the $ab$-plane SPCS on CsV$_3$Sb$_5$ where the sensing current flows parallel to the crystallographic $ab$ plane of CsV$_3$Sb$_5$.  Our previous work has revealed a strongly anisotropic superconducting gap with a pronounced gap minima of 0.1 meV under ambient pressure \cite{ZhaoYQCPB}. Here, we investigate the pressure-dependent PCAR spectra of $ab$-plane CsV$_3$Sb$_5$/Ag junctions across a pressure range of 0.2–1.2 GPa, as summarized in Fig. 5. At a low pressure of 0.18 GPa, the normalized $ab$-plane Andreev conductance spectrum exhibits a prominent broad  in-gap zero-bias conductance peak (ZBCP) with a half-height-width (HHW) of 0.98 meV. This spectral feature is consistent with that observed at ambient pressure \cite{ZhaoYQCPB}. Upon increasing $P$ to 0.42 GPa [not shown here], the PCAR conductance remains essentially unchanged.  With $P$ up to 0.55 GPa [Fig. 5(b)] and 0.65 GPa approaching to $P_1$ [Fig. 5(c)], the broad ZBCP persists, while the corresponding HHW slightly expands to 1.12 meV at 0.65 GPa and 2.0 meV at 0.98 GPa [Fig. 5(e)].  Generally, the broadening of the in-gap HHW is consistent with a reduction of the gap anisotropy, as further supported by the systematic anisotropic-BTK fitting.
	
The pressure-driven evolution of gap anisotropy in $ab$-plane PCAR spectra is further validated via semi-quantitative fitting using a modified BTK model. Considering the six-fold rotational symmetry of the kagome lattice, we phenomenologically parameterize the angular dependence of the gap using the lowest-order six-fold harmonic symmetry  $\Delta(\theta) = \Delta [1+r \cos(6\theta)]$ to describe the angular-dependent gap variation \cite{RoppNC2023}. Here, $\theta$ denotes the in-plane azimuthal angle, and the parameter $r$ characterizes the degree of gap anisotropy. For $0<r<1$, the gap remains nodeless but becomes increasingly anisotropic with increasing $r$, whereas $r=1$ corresponds to the limiting case with gap nodes. This anisotropic $s$-wave BTK model reproduces the low-$T$ normalized conductance $G(V)/G_N$ with excellent accuracy. At $P=0.18$ GPa, the optimal fitting yields a gap amplitude of $\Delta\simeq 0.42$ meV and an anisotropy of $r\simeq 0.80$, confirming a highly anisotropic superconducting gap at low-$P$s.
	
The gradual transition from a highly anisotropic gap toward a nearly isotropic gap under increasing $P$ resembles the gap evolution induced by chemical doping in Ti-substituted CsV$_3$Sb$_5$ systems. Previous STM studies have revealed V-shaped gap features for lightly doped CsV$_{3-x}$Ti$_x$Sb$_5$ ($x=0, 0.03$) and U-shaped, nearly isotropic gap profiles for heavily doped samples ($x=0.10, 0.25, 0.40$) \cite{GaoHJVUgap}. This similarity suggests that chemical substitution and hydrostatic pressure can produce qualitatively similar changes in the superconducting gap anisotropy, although the microscopic mechanisms underlying these two tuning approaches may be distinct.
	
It is interesting to note that as $P$ approaches $P_1$, the $ab$-plane PCAR conductance curves exhibit a pronounced peak-to-dip feature emerging near the superconducting gap edge, as indicated by the arrows in Figs. 5(c)–(f).  The emergence of the peak-dip-hump structure is suggestive of enhanced coupling between superconducting quasiparticles and a bosonic excitation \cite{GaoHJmode}. Such spectral features have been discussed in the context of strong-coupling superconductivity, where the relevant bosonic mode may originate from phononic or electronic collective excitations. When the system is pressure-tuned toward $P_1$, The peak-dip structure becomes pronounced upon entering the pressure regime above $P_1$.  This observation further supports the enhanced superconducting coupling inferred from the gap-ratio analysis and suggests that the coupling between superconducting quasiparticles and a bosonic excitation becomes particularly pronounced near $P_1$.

\textbf{Conclusion:} We have carried out pressure-dependent point-contact Andreev reflection spectroscopy measurements on both $c$-axis and $ab$-plane CsV$_3$Sb$_5$/Ag junctions. The highly reproducible, high-quality soft point-contact spectroscopy allows us to probe the superconducting order parameter of the kagome superconductor CsV$_3$Sb$_5$ under hydrostatic compression. Our spectroscopic results provide strong evidence for two distinct $s$-wave superconducting gaps along the $c$ axis and a strongly anisotropic gap within the $ab$ plane. For the $c$-axis junctions, quantitative analysis using a two-band Blonder–Tinkham–Klapwijk (BTK) model yields a large gap satisfying $2\Delta_l/k_BT_c > 3.52$ and a small gap with $2\Delta_s/k_BT_c < 3.52$. By introducing a spectral-weight-averaged gap $2\Delta_\mathrm{avg}$, we find that the corresponding coupling ratio $2\Delta_\mathrm{avg}/k_BT_c$ is enhanced in the pressure regime between $P_1$ and $P_2$. For the $ab$-plane junctions, the most striking observation is the continuous suppression of the in-plane gap anisotropy under hydrostatic pressure, from $r\sim 0.80$ at low $P$s to $r\sim 0.30$ near 1.2 GPa.  Taken together, these observations uncover the intricate interplay between charge-density-wave (CDW) and superconducting orders in pressurized CsV$_3$Sb$_5$.
	
\textbf{ACKNOMLEDMENTS:} C.R. acknowledges financial support from the Joint Fund of Yunnan Provincial Science and Technology Department (Grant No. 2019FY003008) and the National Natural Science Foundation of China. W. Yang thanks financial support from ``17th Graduate Research Innovation Project'' of Yunnan University (Grant No. KC-252512143). H. Wang acknowledges financial support from the National Natural Science Foundation of China (No. 62164006) and Yunnan Fundamental Reseach Projects (No. 202201AS07008).  Y.Qi would like to acknowledge the National Key R\&D Program of China (Grant No.2018YFA0704300).
	
	$\ddag$ Corresponding author, Email: cren@ynu.edu.cn


\begin{thebibliography}{99}
		
		
\bibitem{flatband1}Tan H X, Liu Y, Wang Z and Yan B 2021
\textit{Charge Density Waves and Electronic Properties of Superconducting Kagome Metals}
\href{https://doi.org/10.1103/PhysRevLett.127.046401}
{{\it Phys. Rev. Lett.} {\bf 127} 046401}
		
\bibitem{flatband2}Nunes L H C M and Morais Smith C 2020
\textit{Flat-band superconductivity for tight-binding electrons on a square-octagon lattice}
\href{https://doi.org/10.1103/PhysRevB.101.224514}
{{\it Phys. Rev. B} {\bf 101} 224514}
		
\bibitem{vonHove0}Wang W S, Li Z Z, Xiang Y Y and Wang Q H 2013
\textit{Competing electronic orders on kagome lattices at van Hove filling}
\href{https://doi.org/10.1103/PhysRevB.87.115135}
{{\it Phys. Rev. B} {\bf 87} 115135}
		
\bibitem{vonHove1}Kang M, Fang S, Kim J K, Ortiz B R, Ryu S H, Kim J, Yoo J, Sangiovanni G, Di Sante D, Park B G, Jozwiak C, Bostwick A, Rotenberg E, Kaxiras E, Wilson S D, Park J H and Comin R 2022
\textit{Twofold van Hove singularity and origin of charge order in topological kagome superconductor CsV$_3$Sb$_5$}
\href{https://doi.org/10.1038/s41567-021-01451-5}
{{\it Nat. Phys.} {\bf 18} 301-308}
		
\bibitem{vonHove2}Kiesel M L, Platt C and Thomale R 2012
\textit{Sublattice interference in the kagome Hubbard model}
\href{https://doi.org/10.1103/PhysRevB.86.121105}
{{\it Phys. Rev. B} {\bf 86} 121105(R)}
		
\bibitem{vonHove3}Cho S Y, Ma H Y, Xia W, Yang Y C, Liu Z T, Huang Z, Jiang Z C, Lu X L, Liu J S, Liu Z H, Li J, Wang J H, Liu Y, Jia J F, Guo Y F, Liu J P and Shen D W 2021
\textit{Emergence of new van Hove singularities in the charge density wave state of a topological kagome metal RbV$_3$Sb$_5$}
\href{https://doi.org/10.1103/PhysRevLett.127.236401}
{{\it Phys. Rev. Lett.} {\bf 127} 236401}
		
\bibitem{topolg1}Yu S L and Li J X 2012
\textit{Chiral superconducting phase and chiral spin-density-wave phase in a Hubbard model on the kagome lattice}
\href{https://doi.org/10.1103/PhysRevB.85.144402}
{{\it Phys. Rev. B} {\bf 85} 144402}
		
\bibitem{topolg2}Ortiz B R, Teicher S M L, Hu Y, Zuo J L, Sarte P M, Schueller E C, Abeykoon A M M, Krogstad M J, Rosenkranz S, Osborn R, Seshadri R, Balents L, He J and Wilson S D 2020
\textit{CsV$_3$Sb$_5$: A $\mathbb{Z}_2$ Topological Kagome Metal with a Superconducting Ground State}
\href{https://doi.org/10.1103/PhysRevLett.125.247002}
{{\it Phys. Rev. Lett.} {\bf 125} 247002}
		
\bibitem{OrtizPRM2019}Ortiz B R, Gomes L C, Morey J R, Winiarski M, Bordelon M, Mangum J S, Oswald I W H, Rodriguez-Rivera J A, Neilson J R, Wilson S D, Ertekin E, McQueen T M and Toberer E S 2019
\textit{New kagome prototype materials: discovery of KV$_3$Sb$_5$, RbV$_3$Sb$_5$, and CsV$_3$Sb$_5$}
\href{https://doi.org/10.1103/PhysRevMaterials.3.094407}
{{\it Phys. Rev. Mater.} {\bf 3} 094407}
		
\bibitem{OrtizPRM2021}Jiang K, Wu T, Yin J X, Wang Z, Hasan M Z, Wilson S D, Chen X H and Hu J P 2023
\textit{Kagome superconductors AV$_3$Sb$_5$ (A = K, Rb, Cs)}
\href{https://doi.org/10.1093/nsr/nwac199}
{{\it Natl. Sci. Rev.} {\bf 10} nwac199}
		
\bibitem{LeiHCCPL2021}Yin Q, Tu Z, Gong C, Fu Y, Yan S and Lei H 2021
\textit{Superconductivity and Normal-State Properties of Kagome Metal RbV$_3$Sb$_5$ Single Crystals}
\href{https://doi.org/10.1088/0256-307X/38/3/037403}
{{\it Chin. Phys. Lett.} {\bf 38} 037403}
				
\bibitem{GiantHall1}Guguchia Z, Verezhak J A T, Gawryluk D J, Tsirkin S S, Yin J X, Belopolski I, Zhou H, Simutis G, Zhang S S, Cochran T A, Chang G, Pomjakushina E, Keller L, Skrzeczkowska Z, Wang Q, Lei H C, Khasanov R, Amato A, Jia S, Neupert T, Luetkens H and Hasan M Z 2020
\textit{Tunable anomalous Hall conductivity through volume-wise magnetic competition in a topological kagome magnet}
\href{https://doi.org/10.1038/s41467-020-15161-y}
{{\it Nat. Commun.} {\bf 11} 559}
		
\bibitem{GiantHall2}Yang S, Liu Y, Zheng H, Wang Y, Yang Y, Zhao J, Chen K, Wang C, Guo J, Chen X, Yang Y, Liu Y, Wang S, Li H, Zhang Z and others 2020
\textit{Giant, unconventional anomalous Hall effect in the metallic frustrated magnet candidate, KV$_3$Sb$_5$}
\href{https://doi.org/10.1126/sciadv.aay3873}
{{\it Sci. Adv.} {\bf 6} eaay3873}
		
\bibitem{GiantHall3}Yu F H, Wu T, Wang Z Y, Lei B, Zhuo W Z, Ying J J and Chen X H 2021
\textit{Concurrence of anomalous Hall effect and charge density wave in a superconducting topological kagome metal}
\href{https://doi.org/10.1103/PhysRevB.104.L041103}
{{\it Phys. Rev. B} {\bf 104} L041103}
		
\bibitem{timeRev1}Mielke III C, Das D, Yin J X, Liu H, Gupta R, Jiang Y X, Medarde M, Wu X, Lei H C, Chang J, Dai P, Si Q, Miao H, Thomale R, Neupert T, Shi Y, Khasanov R, Luetkens H and Guguchia Z 2022
\textit{Time-reversal symmetry breaking in the superconducting state of a kagome metal}
\href{https://doi.org/10.1038/s41586-021-04327-z}
{{\it Nature} {\bf 602} 245-250}
		
\bibitem{timeRev2}Deng H, Liu G, Guguchia Z, Yang T, Liu J, Wang Z, Xie Y, Shao S, Ma H, Liège W, Bourdarot F, Yan X Y, Qin H, Mielke III C, Khasanov R, Luetkens H, Wu X, Chang G, Liu J, Christensen M H, Kreisel A, Andersen B M, Huang W, Zhao Y, Bourges P, Yao Y, Dai P and Yin J X 2024
\textit{Chiral kagome superconductivity modulated by charge order}
\href{https://doi.org/10.1038/s41563-024-01995-w}
{{\it Nat. Mater.} {\bf 23} 1639-1644}
		
\bibitem{nemat1}Xu Y, Ni Z, Liu Y, Ortiz B R, Deng Q, Wilson S D, Yan B, Balents L and Wu L 2022
\textit{Fermi surface reconstruction in the kagome superconductor CsV$_3$Sb$_5$}
\href{https://doi.org/10.1038/s41567-022-01627-7}
{{\it Nat. Phys.} {\bf 18} 1470-1476}
		
\bibitem{nemat2}Xiang Y, Li Q, Li Y, Xie W, Yang H, Wang Z, Yao Y and Wen H H 2021
\textit{Nematic superconductivity and its interplay with charge order in the kagome metal KV$_3$Sb$_5$}
\href{https://doi.org/10.1038/s41467-021-26946-6}
{{\it Nat. Commun.} {\bf 12} 6727}

\bibitem{nemat3}Li Y W, Zhu Z Y, Li Q, Ye Y Z, Wang Z W, Yao Y G and Wen H H 2024

\textit{Electronic Nematicity in the Kagome Superconductor CsV$_3$Sb$_5$}
\href{https://doi.org/10.1088/0256-307X/41/12/127301}
{{\it Chin. Phys. Lett.} {\bf 41} 127301}
		
\bibitem{Pressure1}Yu F H, Ma D H, Zhuo W Z, Liu S Q, Wen X K, Lei B, Ying J J and Chen X H 2021
\textit{Pressure-induced reemergence of superconductivity in a charge-density-wave material}
\href{https://doi.org/10.1038/s41467-021-23850-5}
{{\it Nat. Commun.} {\bf 12} 3645}
		
\bibitem{Pessure2}Chen K Y, Wang N N, Yin Q W, Gu Y H, Jiang K, Tu Z J, Gong C S, Uwatoko Y, Sun J P, Lei H C, Hu J P and Cheng J G 2021
\textit{Double superconducting dome and associated charge density wave in the kagome metal KV$_3$Sb$_5$ under high pressure}
\href{https://doi.org/10.1103/PhysRevLett.126.247001}
{{\it Phys. Rev. Lett.} {\bf 126} 247001}
		
\bibitem{Pressure3}Stier F, Haghighirad A A, Garbarino G, Mishra S, Stilkerich N, Chen D, Shekhar C, Lacmann T, Felser C, Ritschel T, Geck J and Le Tacon M 2024
\textit{Pressure-induced electronic phase transitions in the kagome superconductor CsV$_3$Sb$_5$}
\href{https://doi.org/10.1103/PhysRevLett.133.236503}
{{\it Phys. Rev. Lett.} {\bf 133} 236503}
		
\bibitem{TiSub1}Liu Y X, Wang Y, Cai Y Q, Hao Z Y, Ma X M, Wang L, Liu C, Chen J, Zhou L, Wang J H and others 2023
\textit{Doping evolution of superconductivity, charge order, and band topology in hole-doped topological kagome superconductors Cs(V$_{1-x}$Ti$_x$)$_3$Sb$_5$}
\href{https://doi.org/10.1103/PhysRevMaterials.7.064801}
{{\it Phys. Rev. Mater.} {\bf 7} 064801}
		
\bibitem{TiSub2}Yang H T, Huang Z H, Zhang Y H, Zhao Z, Shi J, Luo H L, Zhao L, Qian G, Tan H X, Hu B, Zhu K, Lu Z, Zhang H, Sun J P, Cheng J G, Shen C M, Lin X, Yan B, Zhou X J, Wang Z, Pennycook S J, Chen H, Dong X L, Zhou W and Gao H J 2022
\textit{Titanium doped kagome superconductor CsV$_{3-x}$Ti$_x$Sb$_5$ and two distinct phases}
\href{https://doi.org/10.1016/j.scib.2022.10.015}
{{\it Sci. Bull.} {\bf 67} 2176-2184}
			
\bibitem{SnSub1}Oey Y M, Ortiz B R, Kaboudvand F, Frassineti J, Garcia E, Cong R, Sanna S, Mitrovi\'c V F, Seshadri R and Wilson S D 2022
\textit{Fermi level tuning and double-dome superconductivity in the kagome metal CsV$_3$Sb$_{5-x}$Sn$_x$}
\href{https://doi.org/10.1103/PhysRevMaterials.6.L041801}
{{\it Phys. Rev. Mater.} {\bf 6} L041801}
		
		
\bibitem{SnSub2}Kautzsch L, Oey Y M, Li H, Ren Z, Ortiz B R, Pokharel G, Seshadri R, Ruff J, Kongruengkit T, Harter J W, Wang Z, Zeljkovic I and Wilson S D 2023
\textit{Structural evolution of the kagome superconductor CsV$_3$Sb$_5$ under pressure}
\href{https://doi.org/10.1038/s41535-023-00570-x}
{{\it npj Quantum Mater.} {\bf 8} 37}
		
		
\bibitem{ZiHPRB2024}Zi H, Zhao Y Q, He M C, Long Y J, Zhao L X, Hou X Y, Yang H X, Yang Y F, Shan L, Ren Z A, Li J Q, Hu J P, Chen G F, Xiong P and Ren C 2024
\textit{Observation of nodal superconductivity in kagome metal CsV$_3$Sb$_5$}
\href{https://doi.org/10.1103/PhysRevB.109.064510}
{{\it Phys. Rev. B} {\bf 109} 064510}
		
\bibitem{LuXSC2023}Zhang D T, Chen C F, Yin L C, Huang Y E, Shi F R, Liu Y, Xu X F, Yuan H Q and Lu X 2023
\textit{Superconducting gap evolution of kagome metal CsV$_3$Sb$_5$ under pressure}
\href{https://doi.org/10.1007/s11433-022-1979-y}
{{\it Sci. China Phys. Mech. Astron.} {\bf 66} 227411}
		
\bibitem{ChenXHSB2023}Wen X, Yu F H, Gui Z, Zhang Y, Hou X, Shan L, Wu T, Xiang Z, Wang Z, Ying J J and Chen X H 2023
\textit{Emergent superconducting fluctuations in compressed kagome superconductor CsV$_3$Sb$_5$}
\href{https://doi.org/10.1016/j.scib.2023.01.020}
{{\it Sci. Bull.} {\bf 68} 259-265}
		
\bibitem{ItalyPC}Gonnelli R S, Daghero D, Ummarino G A, Stepanov V A, Jun J, Kazakov S M and Karpinski J 2002
\textit{Direct Evidence for Two-Band Superconductivity in MgB$_2$ Single Crystals from Directional Point-Contact Spectroscopy in Magnetic Fields}
\href{https://doi.org/10.1103/PhysRevLett.89.247004}
{{\it Phys. Rev. Lett.} {\bf 89} 247004}
		
\bibitem{BTK2}Blonder G E and Tinkham M 1983
\textit{Metallic to tunneling transition in Cu-Nb point contacts}
\href{https://doi.org/10.1103/PhysRevB.27.112}
{{\it Phys. Rev. B} {\bf 27} 112-118}
		
\bibitem{HeMCPRB2022}He M C, Zi H, Zhan H X, Zhao Y Q, Ren C, Hou X Y, Shan L, Wang Q H, Yin Q W, Tu Z, Gong C, Lei H, Lu Z Y, Wang Q, Qi Y P, Chen G F and Xiong P 2022
\textit{Strong-coupling superconductivity in the kagome metal revealed by soft point-contact spectroscopy}
\href{https://doi.org/10.1103/PhysRevB.106.104510}
{{\it Phys. Rev. B} {\bf 106} 104510}
		
\bibitem{LuXPRB2021}Yin L C, Zhang D T, Chen C F, Ye G, Yu F H, Ortiz B R, Luo S S, Duan W Y, Su H, Ying J J, Wilson S D, Chen X H, Yuan H Q, Song Y and Lu X 2021
\textit{Strain-sensitive superconductivity in the kagome metals KV$_3$Sb$_5$ and CsV$_3$Sb$_5$ probed by point-contact spectroscopy}
\href{https://doi.org/10.1103/PhysRevB.104.174507}
{{\it Phys. Rev. B} {\bf 104} 174507}
		
\bibitem{MgB2}Iavarone M, Karapetrov G, Koshelev A E, Kwok W K, Crabtree G W, Hinks D G, Kang W N, Choi E M, Kim H J, Kim H J and Lee S I 2002
\textit{Two-Band Superconductivity in MgB$_2$}
\href{https://doi.org/10.1103/PhysRevLett.89.187002}
{{\it Phys. Rev. Lett.} {\bf 89} 187002}
		
\bibitem{MgB3}Szab\'{o} P, Samuely P, Ka\v{c}mar\v{c}\'{i}k J, Klein T, Marcus J, Fruchart D, Miraglia S, Marcenat C and Jansen A G M 2001
\textit{Evidence for Two Superconducting Energy Gaps in MgB$_2$ by Point-Contact Spectroscopy}
\href{https://doi.org/10.1103/PhysRevLett.87.137005}
{{\it Phys. Rev. Lett.} {\bf 87} 137005}
		
\bibitem{ZhaoYQCPB}Zhao Y Q, Wu Z F, Zuo H Y, Lao W M, Yang W J, Chen Q X, He Y, Wang H, Lv W, Wang Q, Qi Y P, Mu G, Lei H C and Ren C 2026
\textit{Evidence of a strongly anisotropic superconducting gap in kagome superconductor CsV$_3$Sb$_5$: A study of directional point-contact Andreev reflection spectroscopy}
\href{https://doi.org/10.1088/1674-1056/ae27b0}
{{\it Chin. Phys. B} {\bf 35} 057405}
		
\bibitem{RoppNC2023}Roppongi M, Ishihara K, Tanaka Y, Ogawa K, Okada K, Liu S, Mukasa K, Mizukami Y, Uwatoko Y, Grasset R, Konczykowski M, Ortiz B R, Wilson S D, Hashimoto K and Shibauchi T 2023
\textit{Bulk evidence of anisotropic s-wave pairing with no sign change in the kagome superconductor CsV$_3$Sb$_5$}
\href{https://doi.org/10.1038/s41467-023-36273-x}
{{\it Nat. Commun.} {\bf 14} 667}
		
\bibitem{GaoHJVUgap}Yang H T, Huang Z H, Zhang Y H, Zhao Z, Shi J, Luo H L, Zhao L, Qian G, Tan H X, Hu B, Zhu K, Lu Z Y, Zhang H, Sun J P, Cheng J G, Shen C M, Lin X, Yan B, Zhou X J, Wang Z Q, Pennycook S J, Chen H, Dong X L, Zhou W and Gao H J 2022
\textit{Titanium doped kagome superconductor CsV$_{3-x}$Ti$_x$Sb$_5$ and two distinct phases}
\href{https://doi.org/10.1016/j.scib.2022.10.015}
{{\it Sci. Bull.} {\bf 67} 2176-2185}
		
\bibitem{GaoHJmode}Hu B, Chen H, Ye Y, Huang Z, Han X, Zhao Z, Xiao H, Lin X, Yang H, Wang Z and Gao H J 2024, \textit{Evidence of a distinct collective mode inKagome superconductors},
\href{https://doi.org/10.1038/s41467-024-50330-z}
{Nat. Commun.15 6109}
		
\end{thebibliography}
\end{document}